\documentstyle[12pt,epsf]{article}

\newcommand{\beqa}{\begin{eqnarray}}
\newcommand{\eeqa}{\end{eqnarray}}
\newcommand{\beq}{\begin{equation}}
\newcommand{\eeq}{\end{equation}}

\begin{document}
\title{A study of the electronic properties of liquid alkali metals. A
self--consistent approach.}
\author{W. Geertsma$^{1)}$, D. Gonzalez$^{2)}$ and L. H. Gonzalez$^{3)}$ \\
$^{1)}$ Departamento de F\'{i}sica, Universidade Federal do
Esp\'{i}rito
Santo, \\
Av. Fernando Ferrari s/n, 29060-900 Vit\'{o}ria, ES (Brazil)\\
$^{2,3)}$ 
Departemento de Fisica Teorica, 
Universidad de Valladolid, \\
E--47011 Valladolid 
(Spain).}
\maketitle

\begin{abstract}
We   study  the    electronic    properties   (density of  states,
conductivity   and  thermopower)  of  some  nearly--free--electron
systems: the liquid  alkali metals   and two liquid alloys, Li--Na
and Na--K.   The  study  has  been   performed       within    the
self--consistent        second    order  Renormalized   Propagator
Perturbation  Expansion (RPE)  for the self--energy.     The input
ionic  pseudopotentials  and static   correlation  functions   are
derived from  the   neutral pseudoatom  method   and  the modified
hypernetted chain theory  of liquids,  respectively.    Reasonable
agreement with experiment is found     for  Na,  K, Rb and  Na--K,
whereas   for    Li   and Cs  and Li--Na   the  agreement  is less
satisfactory
\vfill 
\noindent Submitted to: Braz. J. Phys. \newline
\end{abstract}


\section{INTRODUCTION}

The introduction   of  the    diffraction model \cite{Ziman(1961)}
together   with the  development  of the pseudopotential  concept,
marked the beginning of extensive  calculations of structural  and
electronic       properties          of   liquid            metals
\cite{Shimoji(1977),Ballentine(1975)}.   Based on this concept one
can justify the use  of simple second--order  perturbation  theory
for  weak scattering systems    like  the alkalis.  The most  used
expression in this Nearly--Free  Electron   (NFE)  model    is the
well--known    Ziman  formula  for  the resistivity.     Two   key
ingredients  appear  in    it, namely the static structure  factor
$S(q)$ and  the  screened    pseudopotential $w(q)$.   These   two
quantities   are   not  independent,   since  the  pseudopotential
determines   the  forces   between   the ions,  which   ultimately
determine the  structure of the liquid.  Therefore $  S(q)$ should
be obtained from $w(q)$ and the result used  in the evaluation  of
the electronic  properties, and   from the new electronic  density
one  should  again derive a new pseudopotential.  This is what  we
call  a {\it  self--consistent}   calculation    of the electronic
properties of a liquid metal.

Most of the calculations  up to date  are, with the  exception  of
ab--initio   Molecular Dynamics simulations, not self--consistent,
i.e.  for $S(q)$ one  takes either   one from  an analysis  of the
experimental  structure factor data,  or one obtained from a model
for the    interatomic interactions,  like   hard   spheres,  with
parameters fitted to experimental data.   This model  is then used
as  input either in an   approximate     scheme    to  solve   the
Ornstein--Zernike  equations like the Percus--Yevick or (Modified)
HyperNetted Chain, or in a liquid structure computer   simulation.
Then a pseudopotential  is chosen and the electronic   properties,
for example   the electronic  density    of   states (EDOS),   are
calculated.   There  have  been some   attempts    to study    the
properties in a self--consistent way in a more restricted   sense,
that is, using the  same pseudopotential for the structure factor,
thermodynamic properties   and  conductivity.    However up to the
present day   such a scheme has failed:  it  leads to disagreement
with  either  the   electronic  properties,   or with  the  atomic
properties.

There are  at least   two   important  exceptions in  the previous
works,  which  do  make the calculations   within    the spirit of
self--consistency.      One is the  work of   Jank    and   Hafner
\cite{Jank(1990)},   where molecular   dynamics are performed with
effective pair  potentials  derived from a given  pseudopotential,
and  then  some  ionic configuration    is  selected  to perform a
Linearized--Muffin--Tin-Orbitals electronic calculation.

The other  calculations, which  are self--consistent,      are the
modern {\it ab--initio} molecular  dynamics   (AIMD)  simulations.
Using this  method, we  are   only  aware, in the case   of alkali
metals,   of EDOS  and conductivity calculations for Na and  Rb at
several temperatures \cite {Silvestrelli(1997),Shimojo(1996)}.

In  this   work,  we propose a completely  different method, which
does not rely on simulations, to obtain within  a self--consistent
scheme, the ionic  structure and   electronic  properties   of the
liquid metal.   Moreover,   the theory for the calculations of the
electronic properties   which is the  second--order   Renormalized
Propagator         Perturbation     Expansion      (RPE)     \cite
{Ballentine(1975),Oosten(1985)},      satisfies    the  so--called
Generalized Optical Theorem  (GOT), which provides a  second level
of self--consistency  in the theory,    namely between the EDOS on
the one hand and the conductivity on the  other hand.  This method
is much  faster  and much less  demanding  than  AIMD simulations,
but, as we will show below, achieves the  same  level of agreement
with experiment  for weak scattering  systems,    like  the alkali
metals, when pseudopotentials of similar quality are used.

A similar RPE--type  scheme to calculate the electronic  structure
has been  applied before  but only  in a non self--consistent  way
(see            for                 a                   discussion
\cite{Shimoji(1977),Ballentine(1975),Oosten(1985)}):  the integral
equation  for the self energy was approximated  by neglecting  the
self    energy   in    the  right    hand     side    of       eq.
\ref{equation:self--energy} below.

The bare pseudopotentials used in this work were  constructed from
first principles  using the Neutral PseudoAtom  (NPA)  method: one
solves a  system  of   one alkali   atom immersed   in an electron
jellium in the Local Density Approximation (LDA). Linear  Response
Theory   (LRT)    was then  applied   to  obtain   the    screened
pseudopotential  $w(q)$   on the one  hand and  the effective pair
potential       on the  other    hand; for details     see   \cite
{Gonzalez(1993),Gonzalez(1996)}.   The structure factor $S(q)$ was
obtained  from the pair  potential  using the Modified HyperNetted
Chain  (MHNC)  theory  of liquids.  This  combination  has already
been applied successfully to the study  of the static  and dynamic
structure  factors of the liquid  alkalis as well as the Na--K and
Li--Na  liquid alloys  \cite {Gonzalez(1993),Gonzalez(1996)},  the
systems studied in this paper.

\section{THEORY}

We give here the essentials  of the theory, further details can be
found    in  \cite{Oosten(1985)}.      Atomic   units with $2m=1$,
$\hbar=1$, $e^2=1$ are used throughout.

We consider  a system of $N$ ions with their corresponding valence
electrons in  a volume $\Omega $. Each valence electron moves in a
self--consistent    potential due to  the ion cores and the  other
valence electrons.   Therefore,  we can write the single--particle
Hamiltonian as: ${\cal H}={\cal  H}_{0}+   {\cal V}$, where ${\cal
H}_{0}$ is the kinetic    energy  operator  and  ${\cal V} (  {\bf
r})=\sum_{i}w(|{\bf    r}-{\bf    R}_{i}|)$     is   an  effective
one--electron   potential which is taken to   be      the   sum of
spherically symmetric screened  local pseudopotentials centered on
each ion. The central quantity in this  study is  the one-electron
Green function,  $G({\bf  k},E)$.  The EDOS per  atom, unit energy
and unit volume is given by 
\begin{equation}
n(E)=-\frac{1}{\pi }\int \frac{d{\bf k}}{(2\pi )^{3}}{\rm Im} G({\bf k},E)
\label{equation:EDOS}
\end{equation}
Moreover in terms  of the so--called   self--energy, $\Sigma ({\bf
k},E)$, the Green function can also be expressed as
\begin{equation}
G({\bf k},E)=(E-k^{2}-\Sigma ({\bf k},E))^{-1}
\label{equation:Green function} 
\end{equation}
Within the second--order     RPE   \cite{Oosten(1985)},        the
self--energy is approximated by 
\begin{equation}
\Sigma ({\bf k},E)=\rho \int \frac{d{\bf k}^{\prime }}{(2\pi )^{3}}w^{2}(|
{\bf k}-{\bf k}^{\prime }|)S(|{\bf k}-{\bf k}^{\prime }|)G({\bf k}^{\prime
},E),
\label{equation:self--energy} 
\end{equation}
where a constant term $\rho w(q=0)$ has been dropped  using  it as
the energy origin, and $\rho $ denotes the atomic number  density.
This expression  for the self  energy is easily extended to liquid
alloys   by   the   substitution        of    $w^{2}(q)S(q)$    by
$\sum_{i,j}(x_{i}x_{j})^{1/2}w_{i}(q)w_{j}(q)S_{ij}(q)$,     where
$x_{i}$  and stands for the concentration and $w_{i}(q)$   for the
screened   pseudopotential     of the   $i$th--component,  whereas
$S_{ij}(q)$ are  the Ashcroft--Langreth  partial static  structure
factors.

Substituting   the relation (\ref{equation:Green  function})  into
(\ref {equation:self--energy}),  an integral equation  is obtained
for the  self--energy,  which is solved for each energy, obtaining
$\Sigma ({\bf k},E)$,  $G({\bf k},E)$, and $n(E)$ for  that energy
from equation (\ref{equation:EDOS}).

The static  electrical     conductivity       is   given    by the
Kubo--Greenwood equation 
\begin{equation}
\sigma =\int dE(-\partial f_{FD}(E)/\partial E)\sigma (E)
\end{equation}
where $f_{FD}(E)$    is the Fermi--Dirac distribution function and
$\sigma (E)$ is the contribution  of the electrons   of energy $E$
to the conductivity, which is given by 
\begin{equation}
\sigma (E)=\frac{8}{3\pi }\int \frac{d{\bf k}}{(2\pi )^{3}}\left( K({\bf k}
,E)-\frac{1}{2}{\bf k}\Delta _{{\bf k}}{\rm Re}G({\bf k},E)\right) ,
\label{equation:Kubo-Greenwood} 
\end{equation}
where  $K({\bf  k},E)$  obeys the following  integral     equation
\cite{Bringer(1971)} 
\begin{equation}
\frac{K({\bf k},E)}{\left| G({\bf k},E)\right| ^{2}}=k^{2}+\int \frac{d{\bf k
}^{\prime }}{(2\pi )^{3}}\Lambda ({\bf k},{\bf k}^{\prime },E)\frac{{\bf k}
\dot{{\bf k}}^{\prime }}{k^{\prime }{}^{2}}K({\bf k}^{\prime },E)
\label{equation:self-consistentK} 
\end{equation}
The vertex function, $\Lambda ({\bf k},  {\bf  k}^{\prime   },E)$,
has still   to be specified. A self--consistent calculation of the
EDOS and  the conductivity   requires  the self--energy   and  the
vertex  function to be related by  the generalized optical theorem
(GOT)
\begin{equation}
{\rm Im} \Sigma({\bf k},E) = \int \frac{d{\bf k}^{\prime}}{(2\pi)^3} \Lambda({\bf k
},{\bf k}^{\prime},E) {\rm Im} G({\bf k}^{\prime},E)
\label{equation:GOT} 
\end{equation}
We have chosen 
\begin{equation}
\Lambda({\bf k},{\bf k}^{\prime},E)= \rho w^2(|{\bf k}-{\bf k}^{\prime}|)S(|
{\bf k}-{\bf k}^{\prime}|)
\label{equation:vertex} 
\end{equation}
which clearly satisfies   the    GOT when   the  second--order RPE
expansion   is   made  for   the    self--energy              \ref
{equation:self--energy}.

Finally, once   $\sigma  (E)$ has been  obtained, the thermopower,
$S$, can be  calculated   as  a function of  energy from the  Mott
relation
\begin{equation}
S=-\frac{\pi ^{2}k_{B}^{2}T}{3|e|}\frac{d\,\,\log \sigma (E)}{d\,\,E}
\label{equation:thermopower}
\end{equation}

\section{NUMERICAL DETAILS}

For  the  calculation of the self energy we used  a linear mesh of
about 400  points  with  a cut--off   of the  pseudopotential   at
$4k_F$.  Results   of the various electronic   properties are well
converged   because a twice as high    cut--off does  give results
which   differ only  by about   .1 \%. That  such    a cut--off is
sufficient   also follows  from  the  observation  that the second
contribution to the conductivity  is nearly  independent   of  the
energy.

The rather singular   integrand appearing in the expressions   for
the self energy   and  the   conductivity  was calculated  using a
rational  expansion up to second order in $q$  in each interval of
the $q$--mesh,  and were evaluated  analytically.    Checks  using
various integration schemes  showed  that this scheme gives a very
reliable approximation of these $q$ integrals, with   an estimated
error of .1 \% or less.

The  iteration     was   started from   the  bottom   of the band.
Convergence  of the self energy was obtained within  10 iterations
-- irrespective of its initialization. No mixing  scheme was used.
The  convergence of the integral equation for the conductivity was
rather slow near the bottom of the  band, requiring  some 20 to 50
iterations.  For the DOS as   a  function  of  the energy from the
bottom of  the band to about  $2E_{F}$   we used  about 50 to  100
energy points.  The solution  of  the integral equation  for   the
conductivity showed oscillations as a function of energy  when the
$q$--mesh is taken too coarse.
\section{RESULTS AND DISCUSSION.}
The preceding   formalism has been applied to liquid alkali metals
at thermodynamic  conditions near the melting point as well  as to
liquid Li--Na and Na--K  alloys at the temperatures  $T=725$ K and
$T=373$  K respectively. Calculations were also performed  for the
thermodynamic   states  considered   by  Silvestrelli      et   al
\cite{Silvestrelli(1997)}  for liquid Na and Shimojo    et  al for
liquid Rb \cite{Shimojo(1996)}.

The results  obtained for  the EDOS    of the liquid  alkalis, and
liquid Li--Na and Na--K    alloys     are    shown   in     figure
\ref{figure:EDOS}.   First,  we note that the present results are,
in general,  very  similar to those derived from the free electron
model and only for   Rb and Cs we  see some structure appearing in
the EDOS.  The bottom of the band is only  slightly  shifted below
the free--electron band; this shift  increases   on alloying  from
about 0.003 Ry. for the pure elements  to 0.02 Ry  for the alloys.
We  checked the validity of these  results of the RPE by comparing
it with results  of  the   Quasi--Crystalline  approximation (QCA)
\cite{Ballentine(1975)} and found only small changes in  the total
EDOS with respect to the RPE results.

The  valence  bands of the alkalis  have   been  investigated   by
Indlekofer and Oelhafen \cite{Indlekofer(1990)}  using various UPS
excitation     energies.  There are  two important characteristics
observed   in these valence  band spectra:  (i) the  widths of the
valence bands, {\it i.e.}  the energy difference between the Fermi
level and the bottom of   the valence band, are  narrower than the
widths expected from free--electron  behaviour, by some  15  to 25
\%,  and (ii) the spectra show a triangular   shape, except for Li
where it is parabolic.      The same narrowing is observed in  the
crystalline state  of alkalis  \cite{Jensen(1985)},   and has been
attributed to electron correlation effects not taken  into account
in the Local Density Approximation \cite {Northrup(1987)}.

The triangular shape  has been shown to be  consistent with a free
electron    EDOS, and  is explained  by the   difference  in cross
sections of the valence electrons  with different angular momentum
for excitations in the UPS  regime   \cite{Jank(1990)}. In the RPE
these different angular moment components are not available.

Although  the  present  scheme   is  not able  to account for  the
bandwidth reduction of the EDOS, it must be remarked that  this is
also the case with the  AIMD simulations  reported for liquid   Na
\cite{Silvestrelli(1997)}   and Rb \cite{Shimojo(1996)}.   In both
cases, the  valence  bands show an almost  parabolic  shape and  a
width similar to the free--electron model.

The electrical  conductivity is calculated by solving the integral
equation  \ref{equation:self-consistentK}  for $K({\bf k},E)$  and
using the vertex defined in \ref{equation:vertex}.  The electrical
conductivity as a function of energy behaves in a similar way  for
all the systems  studied. It is small near the bottom of the band,
and then it  increases to a broad maximum around  the Fermi level,
and then  it decreases.   The Fermi level  is in all cases located
before    the  maximum in $\sigma(E)$,  so according  to  relation
\ref{equation:thermopower}   leading      always     to a negative
thermopower     (see table   \ref{table:data}).  The   sign of the
thermopower agrees with the experimental values except  for Li and
Cs.

In table    \ref{table:data}   we  present  the  resistivity    we
calculated  in the  present   RPE scheme together   with     those
predicted by Ziman's formula  using  the same structure factor and
pseudopotential.  The values of the  latter   approach are in most
cases  somewhat   smaller than    the RPE.   The  comparison  with
experimental data  reveals discrepancies for Li and  Cs,  however,
the discrepancies are smaller in the case of alloys.

In  the  case  of Na at   different   temperatures   we find that,
although  the structure   factor is well   reproduced, the  values
obtained for the  conductivity   are always   smaller   than   the
experimental     ones,  the discrepancy  with experiment  becoming
smaller as the temperature increases.   The same behaviour for the
resistivity       is  obtained   by  Silvestrelli    {\it  et  al}
\cite{Silvestrelli(1997)}  in their AIMD studies for Na; moreover,
the discrepancy   with  experiment   is in their  AIMD calculation
larger for  lower temperatures than in   the RPE.    They used the
Topp--Hopfield pseudopotential for Na, which reproduced  correctly
the structure factor  in all thermodynamic states.   Moreover they
performed    computations     with norm--conserving      non-local
pseudopotentials,    even  including  core  corrections,   but  no
improvement   was  obtained for the conductivity  values, compared
with experiment.

For  Rb, the AIMD    of Shimojo {\it  et al} \cite{Shimojo(1996)},
using   a norm--conserving      non--local      Troullier--Martins
pseudopotential, obtained  conductivities   which are much smaller
than the experimental  ones,  whereas our calculations  agree very
well with  experiment. It is  also interesting   to  note that the
structure factor was correctly reproduced both in the AIMD and  in
our calculations.

We also  performed  calculations for  the  atomic   and electronic
properties     using    the Fiolhais--Perdew       pseudopotential
\cite{Fiolhais(1995)}, but did not obtain  improved agreement with
experiment.
\section{CONCLUSIONS.}
The  above comparisons show   that when suitable  pseudopotentials
are used, the   combination  of LRT,    MHNC    and   RPE produces
self--consistent results for the atomic and electronic  properties
of liquid alkali  metals, which agree with modern state of the art
AIMD methods, at least for  alkalis for which results using   both
these  methods  are  available.     The  agreement  of  our atomic
structure  factors with experiment   is excellent.   The agreement
between   theoretical and experimental electronic   properties  is
less  satisfactory. For Li and  Cs,  we  find large  discrepancies
with  experimental  data  for the resistivity and for thermopower
of  Li  and   Cs  for we do   not even  find the  right   sign. No
thermopower data are available from these AIMD simulations.

In   our  calculations we   have  found that  for  the pure alkali
metals, the  deviation of the EDOS from the free electron parabola
is rather    small, and it increases somewhat    on alloying.  The
resistivity   and  thermopower agree   reasonably     well  with
experiment for  Na, K, Rb  and Na--K but is less  satisfactory for
Li and  Cs,  and as a consequence  for the liquid Li--Na alloys in
the Li--rich composition    regime.      It  is known  that    the
pseudopotential     of Li and Cs  has rather     large  non--local
components,  for Li because   $p$ states  are missing in the core,
and for Cs  because  $d$ states are present near the Fermi  level.
Such non--local components of the pseudopotential   do not seem to
contribute   much  to the  atomic   structure,     but could be of
importance for  the  electronic properties.  However, the use of a
non--local  pseudopotential  by Shimojo \cite {Shimojo(1996)} does
not  improve  the AIMD result   for the  conductivity  for Rb when
compared with experiment. Note that  for Rb we find good agreement
with experiment.    Such non--local    components could  alter the
position of the     maximum in $\sigma$ as  a function  of energy.
Introducing relative stronger  electron--ion scattering below  the
Fermi level, which would give a minimum in $\sigma(E)$,  thus also
changing the sign of the thermopower.

The larger bandwidth  calculated  for  the   EDOS   compared  with
experiment has been attributed to  electron correlations not taken
into account   in the LDA approximation.   In our  scheme  we take
these electron--electron  interactions into account  implicitly in
the  screened pseudopotential.    To take these electron--electron
interactions  into account  in  a more  explicit way would require
the introduction of  a self--energy in the r.h.s. of equation \ref
{equation:self--energy},      and  using the    bare electron--ion
potential: in terms  of diagrams it  means that instead of summing
part  of the electron--electron     interactions   in   a screened
electron--ion    potential, one   collects     them   in a  medium
propagator, which describes  the motion of electrons including the
electron--electron interactions between   two scattering events on
bare electron--ion potentials.  This electron self--energy can  be
approximated  for  these NFE  systems    in   the  Random    Phase
Approximation  (see \cite{Hedin(1965)}). The consequence of such a
scheme  for the conductivity is  not an increase  of the effective
electron--ion scattering,  but could  result in a slight change of
the EDOS at the Fermi level.    Most  of the  changes  in the EDOS
should occur below the Fermi level. The  EDOS at  the Fermi  level
of Na is nearly given by the NFE model.

The discrepancy of the electronic    transport   properties   with
experimental results  we  attribute  to  the construction   of the
pseudopotential,   and  not to  basic problems in our theory which
should  be  valid  for   these weak scattering    systems, as also
follows   from the comparison of our  results  for the resistivity
and those obtained from  the Ziman  equation: we find  only  small
corrections to the Ziman results.

The success of  both   the  present   approach   and  the AIMD  in
reproducing  the experimental   electronic properties appears   to
depend very much  on the subtle  details  of the  pseudopotential.
The atomic   structure and  electronic   structure   and transport
properties   are  determined     by    different   parts  of   the
pseudopotential.    Electrons  penetrate more into the core region
and are also more sensitive to angular non--local  components.  So
changes in these contributions  not necessarily    will affect the
atomic structure.

The conclusion  is that there  exists accurate theoretical   tools
that can in principle give  reliable self--consistent  results for
both   the atomic  and  electronic  properties of  liquid (alkali)
metals  provided   adequate pseudopotentials are  used.   However,
considering  the results of the present calculations,  we conclude
that  most of  the pseudopotentials proposed up to now for  alkali
metals, although good enough to describe  the static and   dynamic
atomic structure, are not accurate enough to give good   agreement
also with the   experimental electronic  properties.  Even for Na,
which  has  always  been  considered   as  the simplest   metal to
describe.

In contrast  to the pseudopotentials used in AIMD simulations  our
pseudopotentials    are adjusted to the electron  density   in the
liquid metal. We attribute the failure of our pseudopotentials  to
describe the EDOS to the linear screening  approximation.  Further
checks are currently being undertaken.   Preliminary results for a
pseudopotential  based on  the electronic  density   show   better
agreement  with  experiment for the electronic   properties, while
preserving  the  good  agreement of  the atomic  properties   with
experiment.  We attribute the  discrepancy between experiment  and
theory in case  of  the  electronic   transport    properties   to
so--called nonlocal contributions  to the pseudopotential,   which
in case of Li and Cs can be rather large.

We  have    shown  that fully      self--consistent     ab--initio
calculations,  with    as  only  input the chosen pseudopotential,
using integral equation approximations  for the static and dynamic
atomic  structure  as well as for the  electronic properties  give
results of the  same reliability as AIMD  simulations,   under the
condition  that  the  electron--ion    pseudopotential    has been
constructed well.  The advantage of the present method is that  it
is much  faster  than the  AIMD  simulations  and only    requires
minimal computational  resources,  and  so the  study  of possible
improvements in the construction   of the  pseudopotential  taking
into account non--local components and  implementation of a scheme
bare electron--ion     potentials         together      with   the
electron--electron    self energy   can in principle   easily   be
implemented and  studied.   
\section*{Acknowledgments.}       Work
supported  by the DGES  (Grant  PB95--0720--C0201),   Junta     de
Castilla y Leon (Grant VA63/96) and   the European Community   TMR
contract   ERBFMBICT--950218 and from,  and   a  grant   from CNPq
(300928/97--0) during the final stages of this work.

\pagebreak
\begin{figure}
\caption[figure:1]   {The EDOS as a  function of   energy for  the
liquid alkalis  near the melting point, and for the liquid  alloys
Li--Na at T=725 K and Na--K at T=373 K.} \label{figure:EDOS}
\end{figure}

\pagebreak

\center{
\begin{table}[b]
\caption[table:1]{Calculated resistivity ($\mu\Omega$cm) and thermopower
( $\mu$V/K) for the liquid alkali metals near melting, the liquid alloys
Li--Na at T=725 K and Na--K at T=373 K. } \label{table:data}
\begin{tabular}{|c|c|c|c|c|c|}
\hline
\hline
\hline
  & $\rho$(Kubo) & $\rho$(Ziman) & $\rho$(exp) & S(Kubo) & S(exp) \\
\hline
Li  & 7.3  & 7.0   & 24  &  -3.9  &  +21.7 \\
Na  & 16.3 & 15.8  & 9.6 &  -8.0  &  -7.9  \\
K   & 19.6 & 18.5  & 13.0&  -10.6 &  -14.0 \\
Rb  & 22.0 & 20.8  & 22.0&  -11.2 &  -6.3  \\
Cs  & 14.9 & 13.8  & 36.0&  -11.7 &  +6.4  \\
\hline
Na  &36.2 &36.4   &24.6  & -16.0  &  -13.2  \\
Li$_{0.4}$Na$_{0.6}$   &27.8    &27.5   &30.2   & -13.9	   &  +2.2  \\
Li$_{0.5}$Na$_{0.5}$   &25.9    &25.3   &31.5   & -13.4	   &  +5.1  \\
Li$_{0.6}$Na$_{0.4}$   &23.2    &22.8   &32.2   & -12.6	   &  +8.5  \\
Li$_{0.8}$Na$_{0.2}$   &18.2    &17.8   &33.1   & -11.1	   &  +16.2  \\
Li  & 12.3 &12.0   &33.6   & -9.0  &  +25.2 \\
\hline
K   &21.6  &20.3   &14.9   & -11.5&	-- \\
Na$_{0.3}$K$_{0.7}$&48.4    &48.2   &39.5   & -8.5	   &	-- \\
Na$_{0.5}$K$_{0.5}$&48.4    &49.0   &43.0   & -8.6	   &	--	\\
Na$_{0.8}$K$_{0.2}$&32.6    &32.8   &26.0   & -8.3	   &	--	\\
Na  &16.3  &15.8   &9.6    & -8.0&	-- \\
\hline
\hline
\hline
\end{tabular}
\end{table}
}

\pagebreak
\section*{}

\epsfverbosetrue
\epsfysize=6in
\epsffile[-1 -1 378 744]{Fig_nfe.eps}
\end{document}